\documentclass[manuscript]{aastex63}
\usepackage{graphicx}
\usepackage{supertabular}
\usepackage{longtable}
\usepackage{CJK}
\usepackage{tikz}
\usepackage{threeparttable}
\usepackage{amsmath}
\usepackage[abs]{overpic}
\usetikzlibrary{shapes,arrows}

\received{----}
\revised{----}
\accepted{Apr 4, 2026}
\submitjournal{AJ}

\begin{document}
\begin{CJK*}{UTF8}{gbsn}
\defcitealias{2025AJ....169..250W}{Paper I}

\title{Extinction Distributions in Nearby Star-resolved Galaxies. II. M33}
\author[0000-0003-3860-5286]{Yuxi Wang (王钰溪)}
\affiliation{Department of Astronomy, College of Physics and Electronic Engineering, Qilu Normal University, Jinan 250200, People's Republic of China; \rm{\href{yuxiwang@qlnu.edu.cn}{yuxiwang@qlnu.edu.cn}}}
\affiliation{Shandong Key Laboratory of Space Environment and Exploration Technology, People's Republic of China}

\author[0000-0003-1218-8699]{Yi Ren (任逸)}
\affiliation{Department of Astronomy, College of Physics and Electronic Engineering, Qilu Normal University, Jinan 250200, People's Republic of China; \rm{\href{yuxiwang@qlnu.edu.cn}{yuxiwang@qlnu.edu.cn}}}
\affiliation{Shandong Key Laboratory of Space Environment and Exploration Technology, People's Republic of China}

\author[0000-0003-4195-0195]{Jian Gao (高健)}
\affiliation{Institute for Frontiers in Astronomy and Astrophysics, Beijing Normal University, Beijing 102206, People's Republic of China}
\affiliation{School of Physics and Astronomy, Beijing Normal University, Beijing 100875, People's Republic of China}

\author[0000-0003-2472-4903]{Bingqiu Chen (陈丙秋)}
\affiliation{South-Western Institute for Astronomy Research, Yunnan University, Kunming, 650500, People's Republic of China}

\author[0009-0001-5020-4269]{Ying Li (李颖)}
\affiliation{Department of Astronomy, College of Physics and Electronic Engineering, Qilu Normal University, Jinan 250200, People's Republic of China; \rm{\href{yuxiwang@qlnu.edu.cn}{yuxiwang@qlnu.edu.cn}}}




\begin{abstract}
Extinction maps are essential for tracing interstellar dust and enabling accurate stellar population studies in galaxies.
Here, a high-resolution extinction distribution of nearby galaxy M33 is constructed by fitting multiband color indexes of the individually resolved red giant branch (RGB) stars from the Panchromatic Hubble Andromeda Treasury: Triangulum Extended Region (PHATTER) survey.
Achieving an angular resolution of approximately 6$^{\prime\prime}$ ($\sim$ 24.4 pc), the extinction map reveals the intricate and heterogeneous distribution of dust throughout the entire disk of M33, with distinct delineation of spiral arms, inter-arm regions, and compact dust clouds.
In addition, it exhibits strong spatial correspondence with the distributions of total hydrogen, H I, and CO, underscoring the reliability of the extinction map for tracing both diffuse and dense components of the interstellar medium. 
The derived $V$-band extinction reaches up to 2.5 mag per pixel, with a mean value of about 1.05 mag.
Beyond providing new insights into the dust structure of M33, the extinction map offers a robust foundation for accurate extinction corrections and will support future studies, including upcoming observations with the Chinese Space Station Telescope.
\end{abstract}

\keywords{dust, extinction; Triangulum Galaxy}


\section{Introduction} \label{sec:intro}

Interstellar extinction refers to the absorption and scattering of starlight by interstellar dust.
Mapping the extinction distributions in galaxies not only enables extinction correction for observations, but also plays a crucial role in determining the physical properties of galaxies \citep{2020ARA&A..58..529S}.
In particular, constructing extinction maps for nearby galaxies is of great practical significance \citep{2018ARA&A..56..673G,2022HabT.........1G}.
On one hand, the diverse interstellar environments in nearby galaxies allow for detailed studies of how dust properties vary with different interstellar and galactic environments.
On the other hand, insights gained from nearby spatially resolved systems can be applied to infer dust distributions and properties in more distant, unresolved galaxies.

In recent years, an increasing number of extinction maps have been constructed for the Large Magellanic Cloud (LMC; e.g., \citealt{2021ApJS..252...23S, 2022MNRAS.511.1317C, 2022MNRAS.516..824B}), the Small Magellanic Cloud (SMC; e.g., \citealt{2019A&A...628A..51J, 2020MNRAS.499..993B, 2021ApJ...910..121N}), and the Andromeda Galaxy (M31; \citealt{2014ApJ...780..172D, 2015ApJ...814....3D, 2016MNRAS.459.2262D, 2025AJ....169..250W}).
These maps reveal the spatial distribution of dust, and serve as powerful tools for accurately probing physical properties and processes within galaxies.
However, for most other nearby galaxies, such extinction maps remain unavailable. 
Consequently, studies often adopt uniform extinction corrections, which can introduce substantial uncertainties into analyses in galaxies.
The lack of resolved extinction information thus limits the ability to fully understand the nature and evolution of these galaxies.

The Triangulum Galaxy (M33) is the third largest member of the Local Group, following M31 and the Milky Way (MW).
The proximity ($\approx 840$ kpc, \citealt{1991ApJ...372..455F}) and nearly face-on orientation make M33 an outstanding laboratory for investigating the structure and evolution of late-type spiral galaxies \citep{1973ugcg.book.....N}.
Furthermore, the nuclear region of M33 is often regarded as an ideal site for studying starburst galaxies (e.g., \citealt{1999ApJ...519..165G}), owing to the active star formation, compact size, and accessibility with current facilities.
Accordingly, studies in M33 offer valuable perspectives on physical processes driving activity in both disk and starburst galaxies.
However, spatially resolved extinction information is essential for accurately correcting stellar photometry and interpreting the intrinsic properties of stars and star-forming regions. 
The construction of an extinction map for M33 is therefore urgently needed to enable precise studies of the internal structure, stellar populations, star formation, among other properties and processes.

With the improvement of the observational capabilities (e.g., wide-field near-IR depth of the Wide Field Camera on the United Kingdom Infra-Red Telescope, \citealt{2013ASSP...37..229I}; sub-arcsecond imaging from Hubble Space Telescope Wide Field Camera 3 and Advanced Camera for Surveys, \citealt{2014ApJS..215....9W, 2021ApJS..253...53W}), an increasing number of member stars and the photometric data in nearby galaxies have been obtained, offering new prospects to map the extinction distributions with unprecedented detail.
By fitting the spectral energy distributions (SEDs) from optical to near-IR for individual member stars from \citet{2021ApJ...907...18R}, \citet[\citetalias{2025AJ....169..250W} hereafter]{2025AJ....169..250W} constructed an updated extinction map of M31 that covers a larger sky area than previous studies.
With the similar method as \citetalias{2025AJ....169..250W}, we select a sample of red giant branch (RGB) stars from the Panchromatic Hubble Andromeda Treasury: Triangulum Extended Region (PHATTER; \citealt{2021ApJS..253...53W}) survey as extinction tracers in this work, and construct a high-resolution extinction map of M33 for the first time.
Section~\ref{sec:data} describes the observational data, while Section~\ref{sec:method} details the methodology employed in this work. 
The results and related discussions are presented in Section~\ref{sec:re}, and the main conclusions are summarized in Section~\ref{sec:conclusion}.


\section{Data and Sample} \label{sec:data}

We adopt red giants in M33 identified from the PHATTER survey as tracers to construct an extinction map in this work.
First, red giants are widely distributed across the galactic disk, allowing high-resolution extinction maps to be constructed on a large scale.
In addition, red giants are brighter than lower main-sequence stars and, unlike asymptotic giant branch (AGB) stars, are less affected by circumstellar dust, leading to more uniform intrinsic colors (e.g., \citealt{2009ApJ...707...89G, 2013ApJ...776....7G, 2013A&A...550A..42C, 2013ApJ...773...30W, 2014A&A...566A.120S, 2016ApJS..224...23X, 2024ApJ...968L..26L}).
Therefore, red giants offer advantages relative to other broadly distributed old stellar populations.

The PHATTER survey \citep{2021ApJS..253...53W}, covering $\sim$ 14 kpc$^2$ of the sky and extending to 3.5 kpc from the center of M33, provides panchromatic resolved stellar photometry for 22 million stars in the near-UV (NUV; $\lambda_{\rm F275W} = 0.272~\mu$m, $\lambda_{\rm F336W} = 0.336~\mu$m), optical ($\lambda_{\rm F475W} = 0.473~\mu$m, $\lambda_{\rm F814W} = 0.798~\mu$m), and near-IR (NIR; $\lambda_{\rm F110W} = 1.120~\mu$m, $\lambda_{\rm F160W} = 1.528~\mu$m) bands.
For each tracer, we first retain photometric measurements with S/N $>$ 4 in each band, and exclude lower-S/N or missing bands from the subsequent calculations to ensure proper photometric quality.
Following the stellar population selection criteria\footnote{As presented in Table 1 of \citet{2023ApJ...957....3S}, the stellar population selection criteria of RGB stars are as follows:\\
a. F110W $<$ 23.5, $q_{\rm F160W} <21$, where $q \equiv {\rm F160W} - ({\rm F110W}-{\rm F160W}-c_0)\times\frac{A_{\rm F160W}/A_V}{A_{\rm F110W}/A_V-A_{\rm F160W}/A_V}$, originally described in \citet{2015ApJ...814....3D}, to ensure good photometric quality.\\
b. IR = GST, F275W != GST, to exclude contamination from UV-bright helium-burning (HeB) stars and retain a purer sample of old red stars.\\
c. Vertices of selection region (F110W$-$F160W, $q_{\rm F160W}$) = (0.6,22), (1.3,22), (1.3,18.7), (0.9,18.7), chosen visually based on clear features in the color-magnitude diagrams}. provided in \citet{2023ApJ...957....3S}, we select an initial sample of RGB stars.
However, as noted in \citet{2023ApJ...957....3S}, the sample of RGB stars is contaminated by a small fraction ($< 5 \%$) of helium-burning stars and asymptotic giant branch stars, which appears as a young ``tail" of the distribution.
As a result, we exclude the bluest 5\% of the initial sample in the F475W$-$F814W color index to obtain a purer RGB sample\footnote{Varying the blue-end threshold from 0\% to 5\% yields only a slight increase in $A_V$ with essentially unchanged histogram shape and map morphology.}, which contains 660,107 tracers in total.

The observed data for each tracer are also obtained from the PHATTER survey.
We adopt the observed color indexes of F336W$-$F475W, F475W$-$F814W, F814W$-$F110W, and F110W$-$F160W to map the dust extinction in M33.
It should be noted that the extinction laws in external galaxies differ from that of the MW.
Therefore, the foreground extinction from the MW must be considered separately when calculating extinction in external galaxies.
In this work, the observed photometric data are first corrected for MW foreground extinction with $E(B-V) \approx 0.06$ mag \citep{2020ApJ...905L..20R} and the average extinction law for diffuse regions in the MW ($R_V = 3.1$, \citealt{2009ApJ...705.1320G,2019ApJ...886..108F,2021ApJ...916...33G, 2022ApJ...930...15D,2024JOSS....9.7023G}; see Table \ref{tab:extinction_law} for details), following \citet{2022ApJS..260...41W}, before calculating extinction.

\begin{deluxetable}{cccc}
    \tablecaption{General extinction predicitons for M33 and MW in multiple bands from optical to IR. \label{tab:extinction_law}}
		\tablehead{	
		\colhead{Band} & \colhead{\hspace{2cm}$\lambda_{\rm eff}$$^a$} & \colhead{\hspace{2cm}$A_{\lambda}/A_V$ (M33)} & \colhead{\hspace{2cm}$A_{\lambda}/A_V$ (MW)} \\
		\colhead{} & \colhead{\hspace{2cm}$(\mu {\rm m})$} & \colhead{\hspace{2cm}\citet{2022ApJS..260...41W}} & \colhead{\hspace{2cm}\citet{2023ApJ...950...86G}, $R_V = 3.1$}
		}
	\startdata	                
	F336W & \hspace{2cm}0.336 & \hspace{2cm}1.709 & \hspace{2cm}1.643\\
	F475W & \hspace{2cm}0.479 & \hspace{2cm}1.169 & \hspace{2cm}1.183\\
	F814W & \hspace{2cm}0.806 & \hspace{2cm}0.531 & \hspace{2cm}0.574\\
	F110W & \hspace{2cm}1.162 & \hspace{2cm}0.380 & \hspace{2cm}0.299\\
	F160W & \hspace{2cm}1.539 & \hspace{2cm}0.246 & \hspace{2cm}0.186
\enddata
	\tablecomments{$^a$ Effective wavelengths of multiple bands used in this work refer to the SVO Filter Profile Service (http://svo2.cab.inta-csic.es/theory/fps/, \citealt{2012ivoa.rept.1015R}).}
	\end{deluxetable}


\section{Method} \label{sec:method}

In this work, the extinction map in M33 is constructed by fitting multiband color indexes of the individual stars.
This approach has been successfully applied in previous studies to derive extinction distributions in the MW (e.g., \citealt{2012ApJ...757..166B, 2014MNRAS.443.1192C, 2021ApJ...906...47G}), Magellanic Clouds (e.g., \citealt{2022MNRAS.511.1317C}), and M31 (\citetalias{2025AJ....169..250W}).
The observed color of a given star in bands of $\lambda_1$ and $\lambda_2$ is a combination of the intrinsic color $(\lambda_1 - \lambda_2)_0$ and color excess (i.e., subtraction of extinctions, $A_{\lambda_1} - A_{\lambda_2}$): 
\begin{equation}
    (\lambda_1 - \lambda_2) = (\lambda_1 - \lambda_2)_0 + (A_{\lambda_1} - A_{\lambda_2}) = (\lambda_1 - \lambda_2)_0 + A_V(A_{\lambda_1}/A_V - A_{\lambda_2}/A_V),
\end{equation}
where $A_\lambda/A_V$ indicates the relative extinction value, commonly used to characterize the wavelength dependence of interstellar extinction.

The values of $A_\lambda/A_V$ for each band in this work are taken from the extinction curves of M33 calculated by \citet{2022ApJS..260...41W}, who adopted a silicate-graphite dust model to simulate the absorption and scattering of starlight by dust, and derived dozens of extinction curves toward different sight lines.
We map the extinction curves toward different sight lines from \citet{2022ApJS..260...41W} into a spatial distribution and assign to each tracer the curve corresponding to the position.
For the regions where extinction curves are unavailable, we adopt the average extinction curve derived by \citet{2022ApJS..260...41W}, of which the $A_\lambda/A_V$ values are listed in Table \ref{tab:extinction_law}.

\begin{figure}[ht!]
    \centering
	\includegraphics[scale = 0.6]{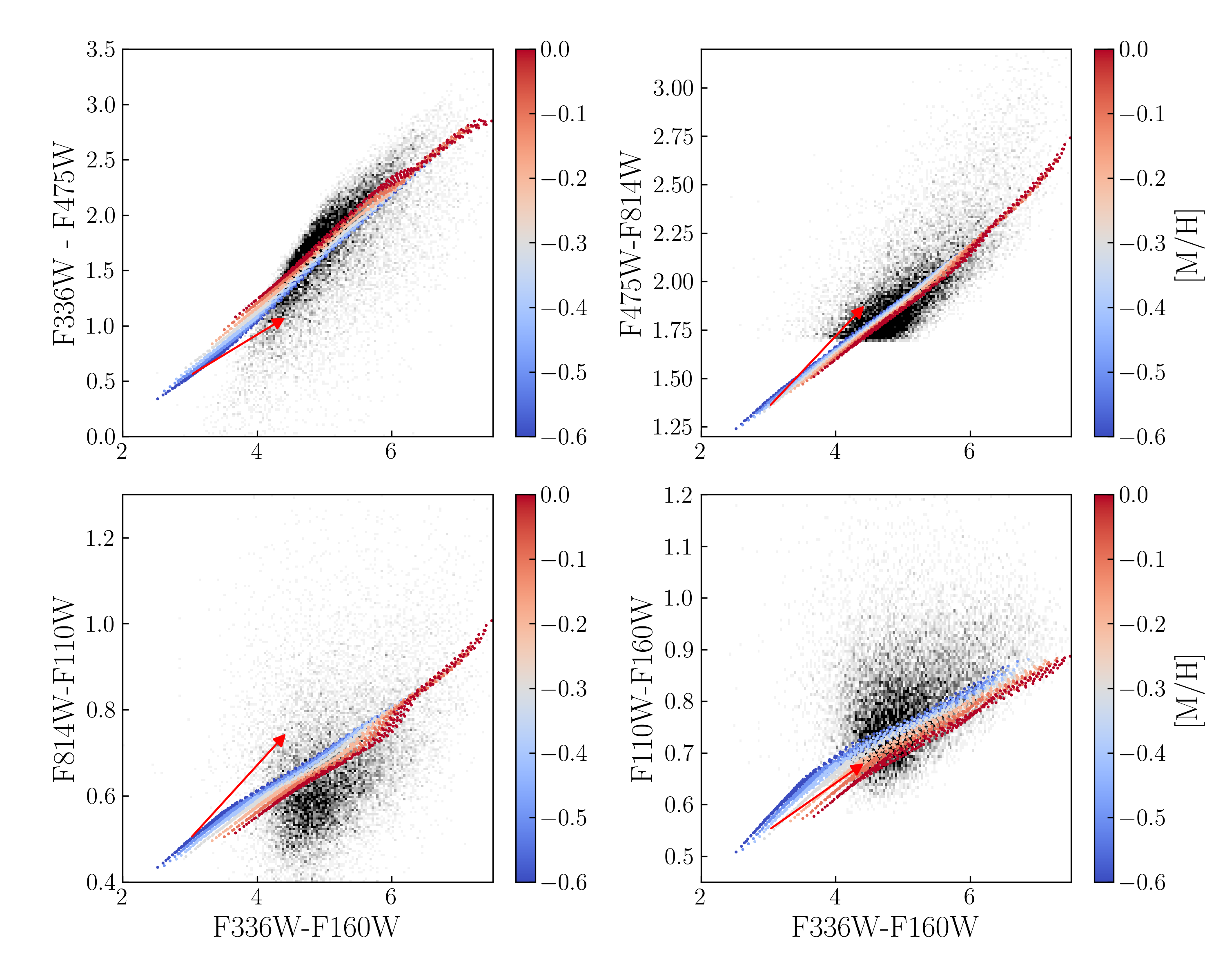}	
	\caption{Color-color diagrams for RGB stars at different metallicities generated with the CMD 3.9 web interface. 
	Gray contours show the observed RGB sample adopted in this work with foreground MW extinction corrected (darker shades indicate higher tracer density), and the red arrows indicate the reddening vector for $A_V = 1$ mag assuming the mean extinction law of M33 \citep{2022ApJS..260...41W}.
	The sharp cutoff at F475W$-$F814W = 1.74 reflects the exclusion of the bluest 5\% of tracers in F475W$-$F814W to reduce contamination from helium-burning and asymptotic giant branch stars, as described in Section \ref{sec:data}.}
	 \label{fig:stellar_loci}
\end{figure}

The intrinsic color indexes $(\lambda_1 - \lambda_2)_0$ for tracers in this work are derived from theoretical stellar loci in multidimensional color space constructed by the \emph{CMD 3.9} web interface\footnote{\url{http://stev.oapd.inaf.it/cgi-bin/cmd}}.
The online tool provides isochrones and synthetic stellar populations based on updated PARSEC evolutionary tracks, enabling us to compute intrinsic magnitudes in the F336W, F475W, F814W, F110W, and F160W bands for a set of model stars.
The input parameters, such as evolutionary tracks, resolution of the thermal pulse cycles, mass loss on the RGB, long period variability, initial mass function and ages, are identical to those in \citetalias{2025AJ....169..250W}, whereas the metallicity grid is newly defined.
For the metallicity, we adopt the [M/H] distribution of M33 derived by \citet{2025AJ....170....2L} to estimate the metallicity of each RGB source in the sample of this work, which results in values spanning approximately $-0.6 \lesssim \mathrm{[M/H]} \lesssim 0.0$
\footnote{Although the radial [M/H] profile of M33 given by \citet{2025AJ....170....2L} can be extrapolated to values as high as $\mathrm{[M/H]}\sim+0.6$ in the innermost regions, we truncate the metallicity distribution at $\mathrm{[M/H]}=0$ in this work. 
On one hand, the metallicities directly derived from the observational data in \citet{2025AJ....170....2L} remain below $\mathrm{[M/H]}=0$.
On the other hand, the conversion from $J-H$ to [M/H] obtained by \citet{2004MNRAS.354..815V} likewise implies that the metallicity in M33 does not exceed $\mathrm{[M/H]}=0$.}.
We therefore compute isochrones at $\mathrm{[M/H]}$ from $-0.6$ to $0.0$ in steps of 0.1 dex and obtain the corresponding intrinsic magnitudes at each metallicity.
As noted by \citet{2023ApJ...957....3S}, the selected RGB sample is primarily populated by evolved stars with masses $\lesssim 2 M_{\odot}$ in the hydrogen shell-burning phase.
Accordingly, we select model red giants (label = 3 in the CMD 3.9 output table) with masses $\lesssim 2 M_{\odot}$ and surface gravity $1 < \log(g) < 3$ for analysis.
Taking (F336W$-$F160W)$_0$ as the independent variable, we present the resulting color-color diagrams for different metallicities in Figure~\ref{fig:stellar_loci}, where gray contours indicate the observed RGB stars selected in Section \ref{sec:data} after correcting for foreground MW extinction, with darker shades indicating higher density\footnote{Although some tracers in individual color-color diagrams do not show the expected shift along the reddening vector relative to the stellar locus, jointly fitting multiple colors enables tracers that deviate in one diagram but follow the reddening trend in others to yield reliable extinction estimates.}.
The median values of each panel, obtained after 3$\sigma$ clipping, are fitted with fifth-order polynomials (e.g., \citealt{2021ApJ...906...47G,2022MNRAS.511.1317C}; \citetalias{2025AJ....169..250W}) to construct the stellar loci.
For each RGB star, we apply the stellar loci corresponding to the estimated metallicity, and the color ranges of the stellar loci adopted in the extinction calculation are summarized in Table \ref{tab:sl_range}.

\begin{deluxetable}{cccc}
    \tablecaption{Color ranges of the stellar loci in (F336W$-$F160W)$_0$ for different metallicities. \label{tab:sl_range}}
		\tablehead{	
		\colhead{[M/H]} & \colhead{\hspace{3cm}(F336W-F160W)$_0, {\rm min}$} & \colhead{\hspace{3cm}(F336W-F160W)$_0, {\rm max}$} \\
		\colhead{} & \colhead{\hspace{3cm}(mag)} & \colhead{\hspace{3cm}(mag)} 
		}
	\startdata	                
	-0.6 & \hspace{3cm}2.6 & \hspace{3cm}6.4 \\
	-0.5 & \hspace{3cm}2.6 & \hspace{3cm}6.6 \\
	-0.4 & \hspace{3cm}3.0 & \hspace{3cm}6.7 \\
	-0.3 & \hspace{3cm}3.0 & \hspace{3cm}6.9 \\
	-0.2 & \hspace{3cm}3.5 & \hspace{3cm}7.0 \\
	-0.1 & \hspace{3cm}4.0 & \hspace{3cm}7.3 \\
	0.0 & \hspace{3cm}4.0 & \hspace{3cm}7.5 
\enddata
	\end{deluxetable}

As mentioned above, the intrinsic color index in Equation (1) can be parameterized by the intrinsic color index between the F336W and F160W bands (F336W$-$F160W)$_0$.
Therefore, there are two parameters, (F336W$-$F160W)$_0$ and $A_V$, that need to be fitted for individual tracers in this work.
Based on the availability of observed color indexes for each source, we divide the 660,107 tracers into four subsamples: (1) 651,115 sources with more than two observed color indexes; (2) 3908 sources with only two observed color index; (3) 4532 sources with only (F336W$-$F160W) and one additional observed color index; and (4) 8992 sources with only one observed color indexes. 
For the last subsample, extinction values cannot be determined. 
For the other three subsamples, the parameters are derived with the same methods as described in \citetalias{2025AJ....169..250W} : For the tracers with more than two observed color indexes, the best-fitting intrinsic color (F336W$-$F160W)$_0$ and extinction value $A_V$ are determined by minimizing the $\chi^2$ function, defined as Eq. (5) in \citetalias{2025AJ....169..250W}. 
For the tracers with two observed color indexes, (F336W$-$F160W)$_0$ and $A_V$ are derived by solving the system of equations given by Eq. (6) in \citetalias{2025AJ....169..250W}.
For the tracers where only the color index (F336W$-$F160W) and one other observed color index are available, the location of the tracer in color-color space is shifted along the reddening vector, and the intersection with the stellar locus indicates the intrinsic color. 
The extinction in the $V$ band is then calculated by comparing the derived intrinsic color with the observed one.
About 2.9\% of the fitting results are excluded from the construction of the extinction map because of $\chi^2$ values exceeding the 3$\sigma$ threshold or unphysical derived parameters (e.g., negative $A_V$ or (F336W$-$F160W)$_0$ outside the ranges listed in Table \ref{tab:sl_range}).
Since the reliability of the results depends on the amount of available photometric data, weights of 0.8 and 0.6 are assigned to the results from the second and third cases, respectively, relative to the first one.

Based on the derived extinction values $A_V$ of individual tracers, we map the extinction distribution in M33 with the HEALPix pixelization scheme \citep{2005ApJ...622..759G}.
The region toward M33 is divided into subfields (HEALPix pixels) with an angular resolution of $\sim$ 6$^{\prime\prime}$ (HEALPix \texttt{nside} = 32,768; $\sim 24.4$ pc), which is sufficient to resolve the main structures while retaining enough tracers per pixel for robust $A_V$ estimates.
In M33, RGB stars are expected to be mixed with the dust (e.g., \citealt{2009A&A...493..453V}), and we cannot accurately determine the relative locations of individual RGB stars with respect to the dust along the sight line.
It is found that approximately 60\% of pixels exhibit $A_V$ distributions consistent with normality, while most of the remainder show a one-sided tail. 
To limit tail-driven bias in the mean, we remove the lowest and highest 20\% of tracers by $A_V$ in each pixel and adopt the weighted mean $A_V$ of the retained tracers as the pixel's extinction.
The extinction distribution constructed in this work thus represents the average extinction of the stars by dust rather than the full line-of-sight dust column, and serves as an effective tool for providing extinction corrections for stellar observations and studies.

\section{Results and discussions} \label{sec:re}

\subsection{Extinction distribution in M33}

\begin{figure}[htbp]
	\centering
	\begin{minipage}{0.32\textwidth}
	  \centering
	  \includegraphics[width=\linewidth,height=0.25\textheight,keepaspectratio]{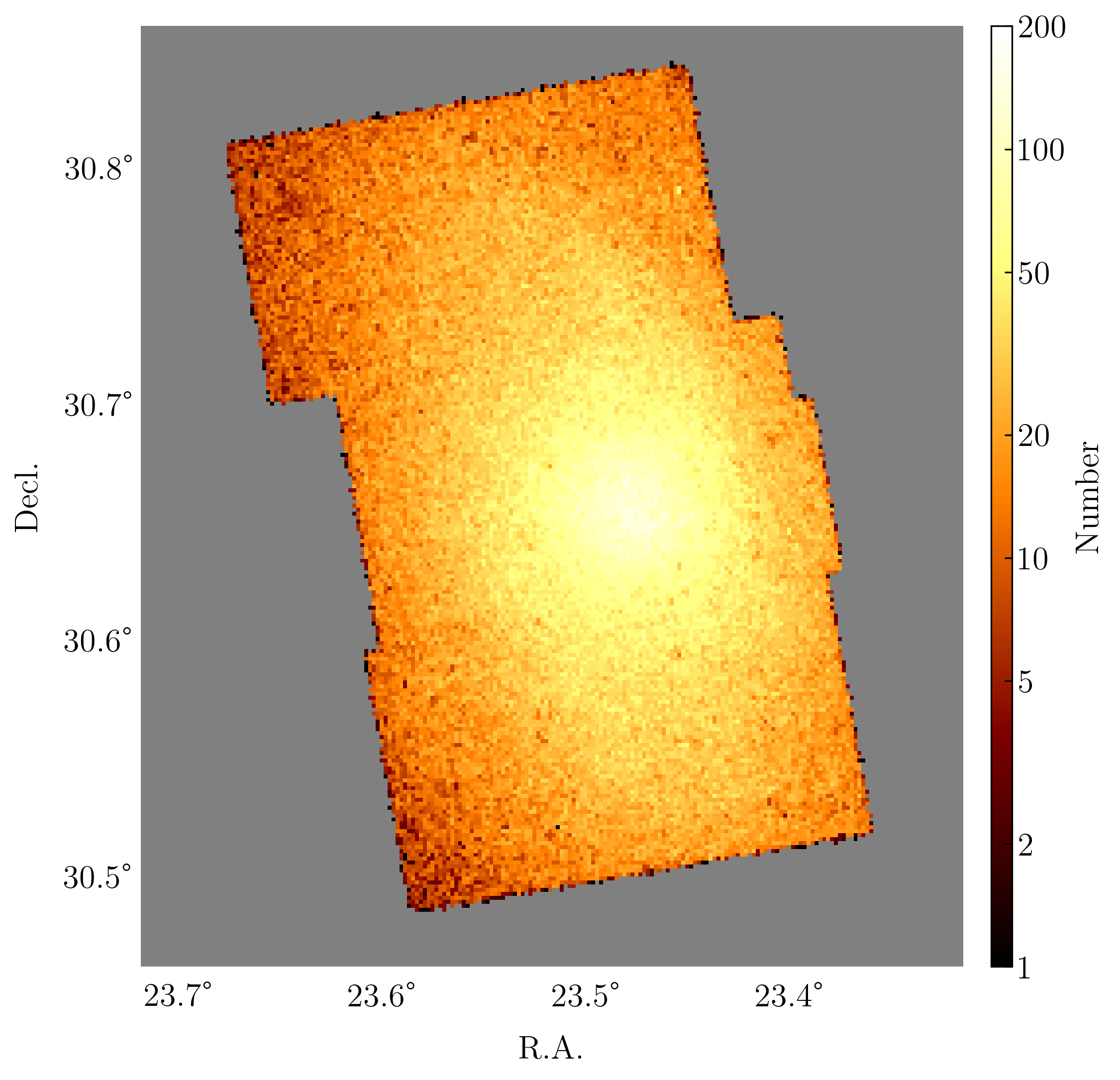}
	\end{minipage}\hfill
	\begin{minipage}{0.32\textwidth}
	  \centering
	  \includegraphics[width=\linewidth,height=0.25\textheight,keepaspectratio]{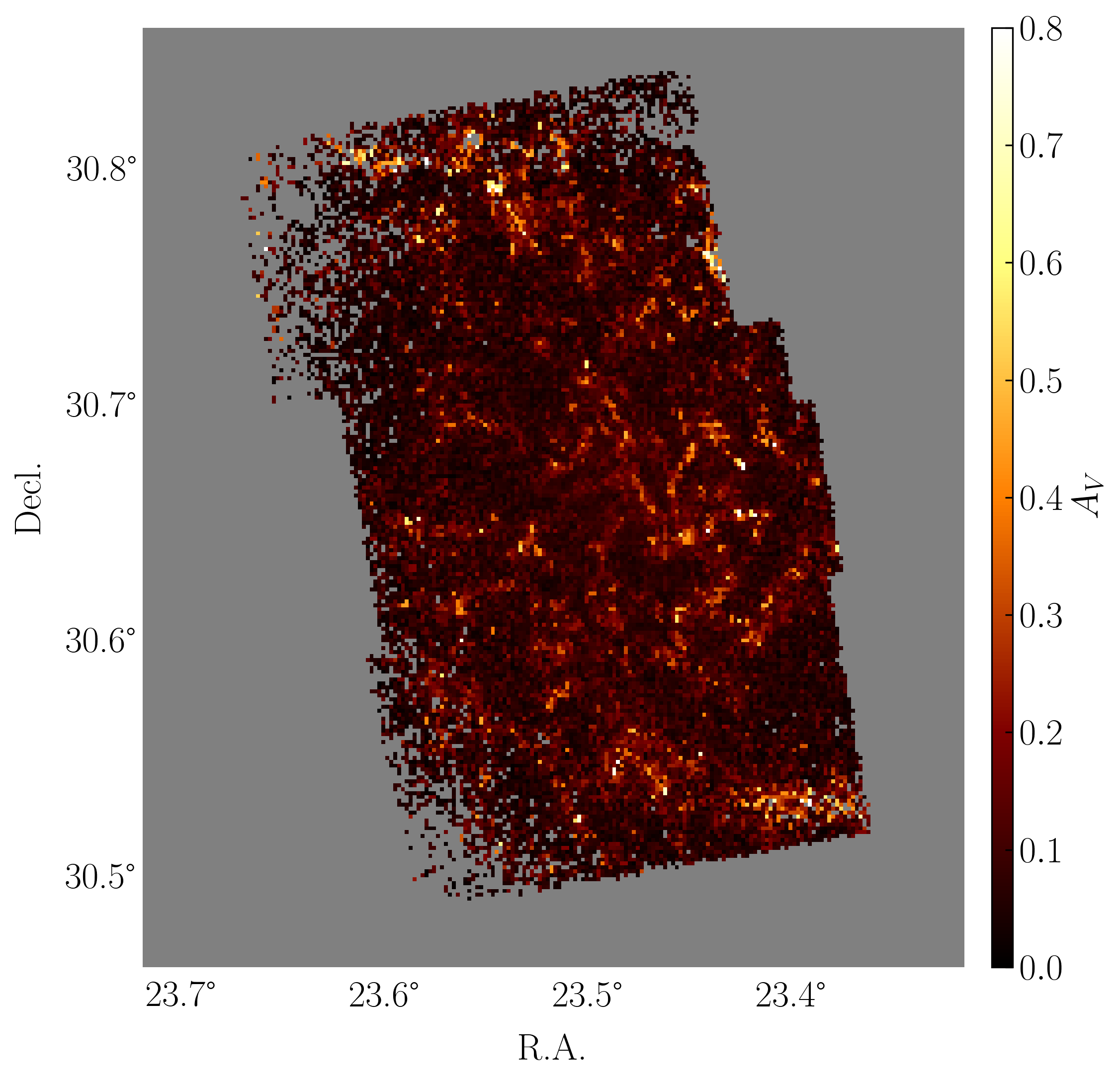}
	\end{minipage}\hfill
	\begin{minipage}{0.3\textwidth}
	  \centering
	  \includegraphics[width=\linewidth,height=0.25\textheight,keepaspectratio]{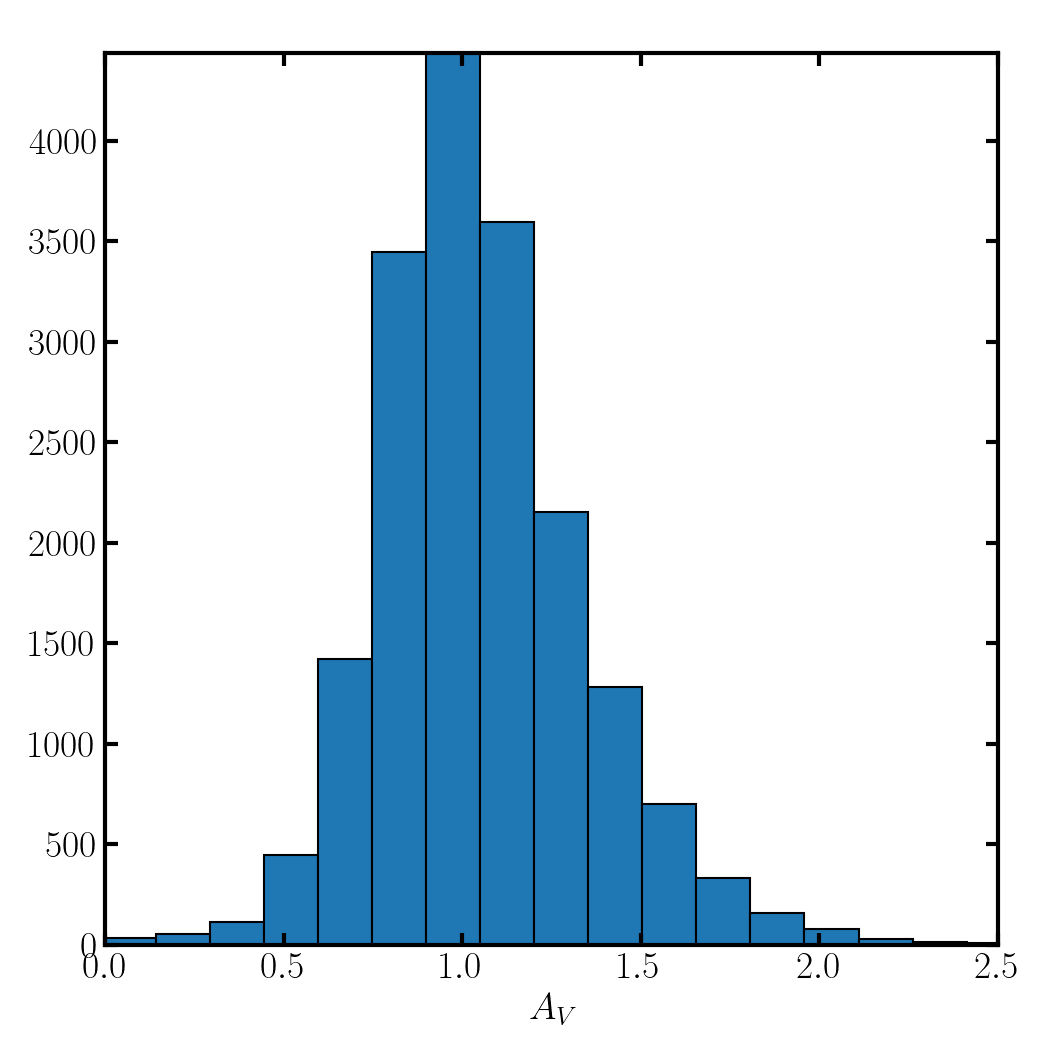}
	\end{minipage}
	\vspace{0.5em} 
	\caption{Left panel: distribution of the stellar counts in each pixel.
	Middle panel: distribution of $A_V$ variance.
	Right panel: histogram for $A_V$ values over all measured pixels.}
	\label{fig:M33_extinction_var}
  \end{figure}

A total of 0.64 million tracers are ultimately used for the construction and analysis of the extinction map.
The tracers provide extinction measurements along individual sight lines, enabling a detailed mapping of the extinction distribution in M33.
The stellar distribution based on the HEALPix pixelization scheme is presented in the left panel of Figure \ref{fig:M33_extinction_var}.
The number of stars per pixel varies significantly across the field, with values ranging from a few up to a maximum of about 180. 
The spatial distribution of tracers defines the effective coverage of the extinction map constructed in this work. 
Higher stellar densities are typically found in the central and brighter regions of M33, while the outskirts and low surface brightness areas contain fewer tracers and thus a lower pixel density.
To ensure statistical reliability in the derived extinction values, we require each pixel to contain more than 10 stars.

  \begin{figure}[htbp]
	\centering
	\begin{minipage}{\textwidth}
	  \centering
	  \includegraphics[width=\linewidth,height=0.75\textheight,keepaspectratio]{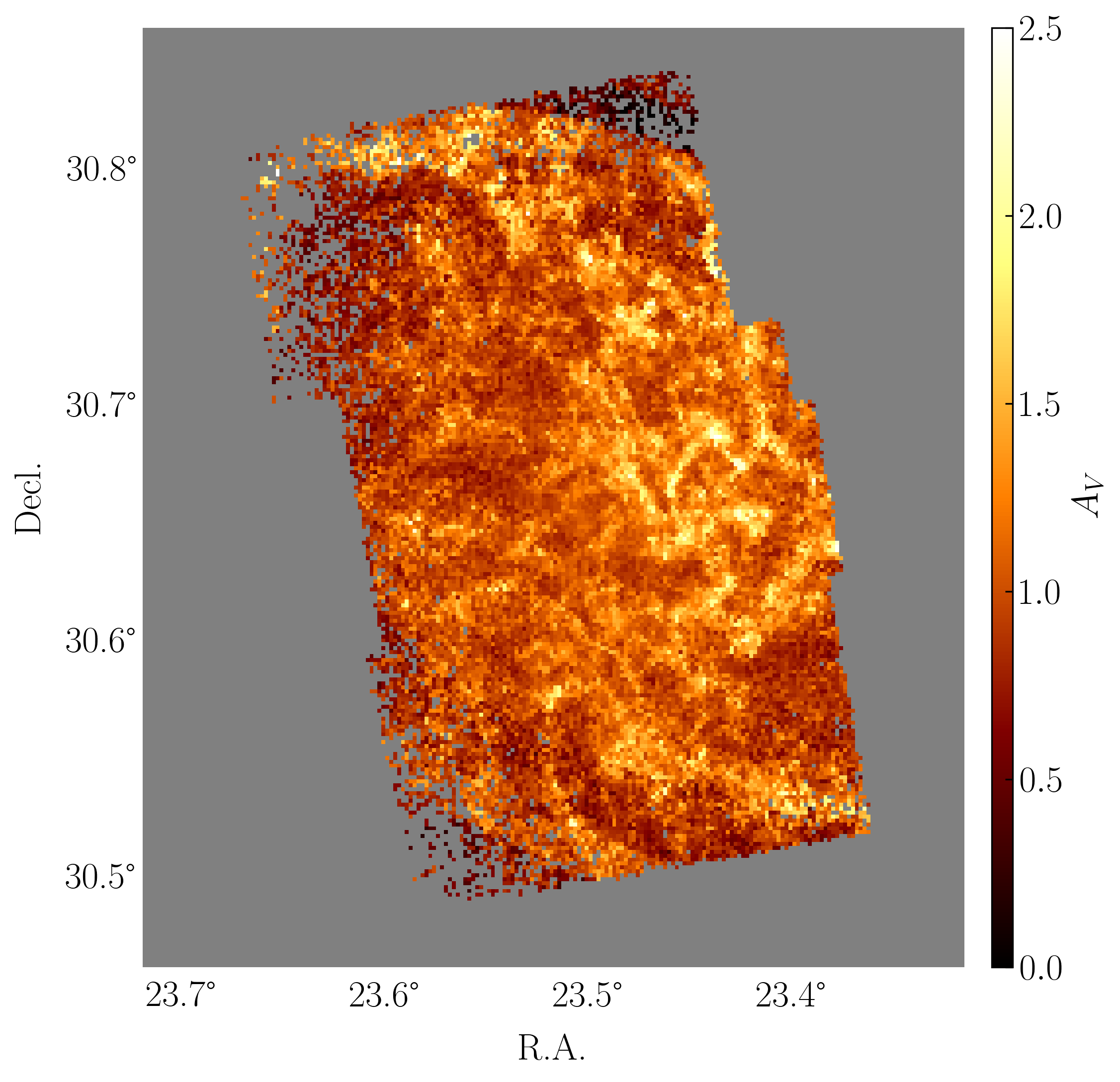}
	\end{minipage}
	\caption{Extinction distribution in M33 with a resolution of approximately 6$^{\prime\prime}$.
	}
	\label{fig:M33_extinction}
  \end{figure}

Figure \ref{fig:M33_extinction} and the middle panel of Figure \ref{fig:M33_extinction_var} present the extinction distribution and the $A_V$ variance distribution, respectively, both at a resolution of approximately 6$^{\prime\prime}$ ($\sim$ 24.4 pc).
Quantitatively, we present a histogram of the extinction values $A_V$ across the extinction map constructed in this work in the right panel of Figure \ref{fig:M33_extinction_var}, and find that the the derived $A_V$ values span from very low extinction up to 2.5 mag.
However, it should be noted that regions with low $A_V$ values may suffer from large uncertainties, since the stellar loci have intrinsic widths and the observed photometry is subject to measurement errors, as discussed in \citet{2015ApJ...814....3D}.
Additionally, based on the intrinsic colors and absolute magnitudes of RGB stars compiled in \citet{2000asqu.book.....C}, together with the PHATTER detection limits, we estimate that the maximum extinction toward an individual sight line constrained in this work is $A_V \approx 4-5$ mag\footnote{We adopt two approaches to estimate the maximum extinction toward individual sight lines.
(1) Brightness method: adopting typical absolute magnitudes of red giants $M_V$ = -0.3 to +0.9 mag \citep{2000asqu.book.....C} and the PHATTER catalog completeness limits $m_{\rm F475W} = 28.5$ mag and $m_{\rm F814W} = 27.5$ mag \citep{2021ApJS..253...53W}, together with the M33 distance modulus $\mu = 24.6$, we infer an upper-limit extinction of $A_V \approx 4$ mag.
(2) Color method: red giants have intrinsic near-IR colors $(J-H)_0  = 0.5 - 0.96$ mag \citep{2000asqu.book.....C}.
Since the effective wavelengths of F110W and F160W are close to those of $J$ and $H$, we approximate (F110W$-$F160W)$_0 \approx 0.5 - 0.96$ mag.
From the lower right panel of Figure \ref{fig:stellar_loci}, the upper limit of the observed color is (F110W$-$F160W) $\approx 1.2$ mag.
Adopting the values of $A_{\rm F110W}/A_V$ and $A_{\rm F160W}/A_V$ listed in Table \ref{tab:extinction_law}, we estimate an upper limit of $A_V \approx 5$ mag.}.

\begin{figure}[t]
	\centering
	\begin{minipage}[t]{0.49\textwidth}
	\centering
	\raisebox{-4mm}{%
\begin{overpic}[height=9.3cm]{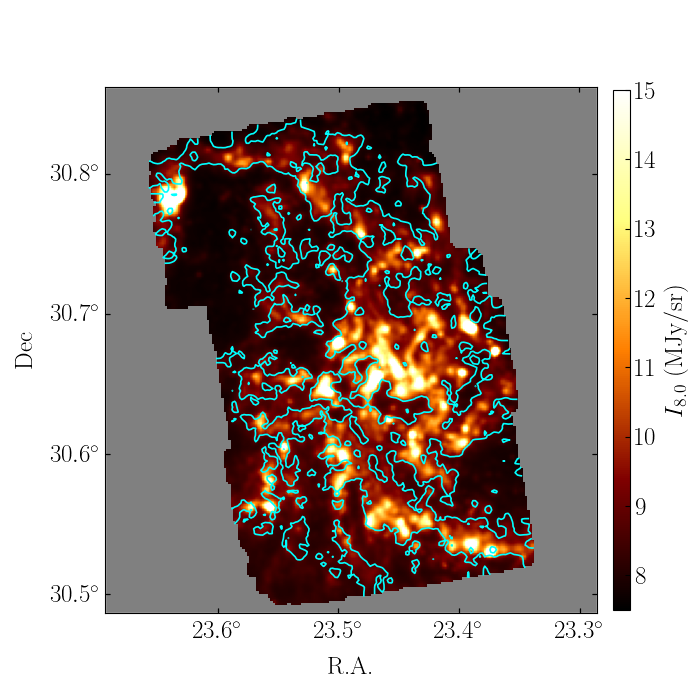}\end{overpic}%
}
	\end{minipage}\hfill
	\begin{minipage}[t]{0.49\textwidth}
	\centering
	\begin{overpic}[height=8.0cm]{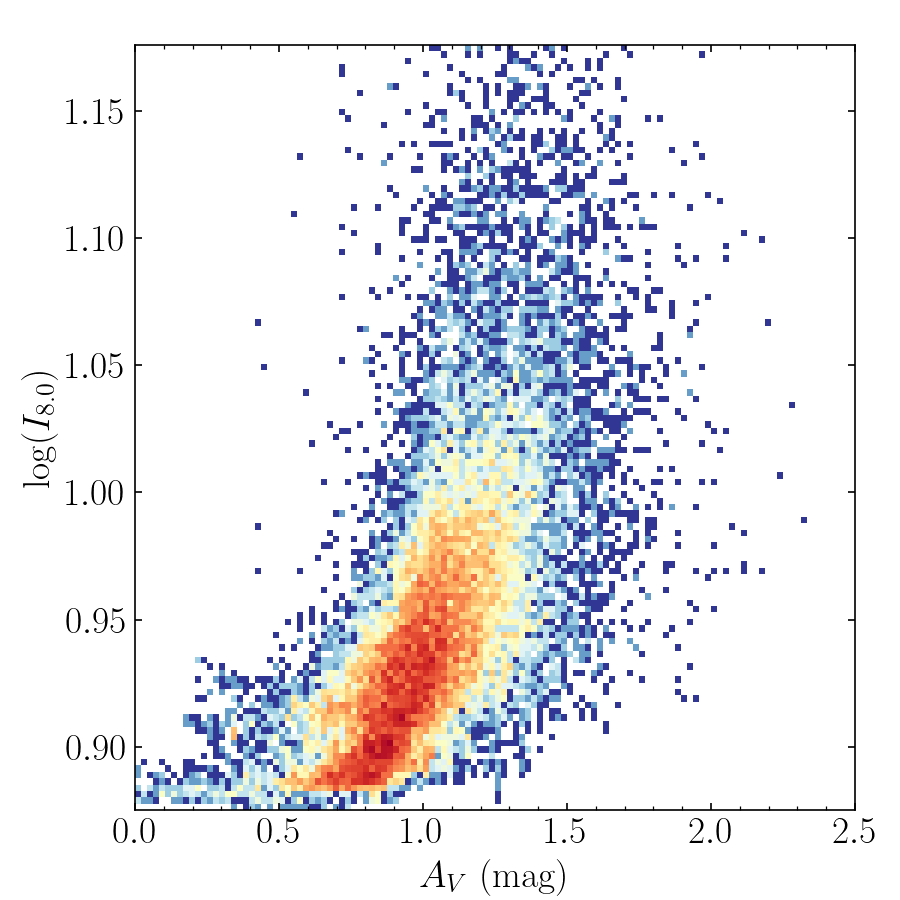}\end{overpic}
	\end{minipage}
	\caption{Comparison between the extinction map constructed in this work and the Spitzer/IRAC 8 $\mu$m emission distribution. 
	Left panel: Spitzer/IRAC 8 $\mu$m map at 12$^{\prime\prime}$ resolution, with cyan contours showing the mean $A_V$ derived in this work. 
	Right panel: pixel-by-pixel comparison, shown as a density plot of $A_V$ versus ${\rm log}(I_{8.0}$), with colors from red to blue indicating high to low point density.}
	\label{fig:com_spitzer}
	\end{figure}

The extinction in Figure \ref{fig:M33_extinction} exhibits substantial spatial variation, showing the complexity and highly patchy nature of the dust distribution across the galaxy.
Notably, the overall morphology of the extinction closely follows the structure seen in the Spitzer/IRAC 8.0~$\mu$m emission map \citep{2009ApJ...703..517D}, which traces the 7.7-micron emission from polycyclic aromatic hydrocarbons (PAHs), as presented in the left panel of Figure \ref{fig:com_spitzer}, indicating a close correspondence between regions of high extinction and infrared-bright regions associated with star formation.
The right panel of Figure \ref{fig:com_spitzer} shows a positive pixel-by-pixel correlation between the $A_V$ derived in this work and $\log(I_{8.0})$, supporting the reliability of the extinction map as a tracer of PAH-rich, dust-dense structures.
Moreover, the regions with elevated extinction are well aligned with the spiral arms observed in multi-band images of M33, indicating that dust plays a crucial role in shaping the observed morphology as seen across various spectral bands. 
On smaller scales, the extinction map reveals a wealth of finer structures, such as filaments, knots, bubbles, and compact clumps, many of which are invisible or less prominent in lower-resolution or single-wavelength observations. 
These small-scale features may be linked to localized star formation regions, molecular clouds, or other dynamic processes in the interstellar medium.
The high resolution of the extinction map constructed in this work enables the investigation of the spatial structure in unprecedented detail, offering new insights into the connections between dust, star formation, and the overall structure of the galaxy.

The extinction map constructed in this work is publicly accessible in Zenodo \citep{wang_2025_17393995}\footnote{doi: \href{https://doi.org/10.5281/zenodo.17393995}{10.5281/zenodo.17393995}}.
Furthermore, the Zenodo repository offers a convenient Python script that can calculate extinction values in the $V$ band for specified coordinates (R.A. and Dec.). 
A usage example is provided as well.

\subsection{Comparison with gas distributions in M33}

\begin{figure}[htbp]
	\centering
  
	\begin{minipage}[t]{0.49\textwidth}
		\centering
		\raisebox{-4mm}{%
	\begin{overpic}[height=8.9cm]{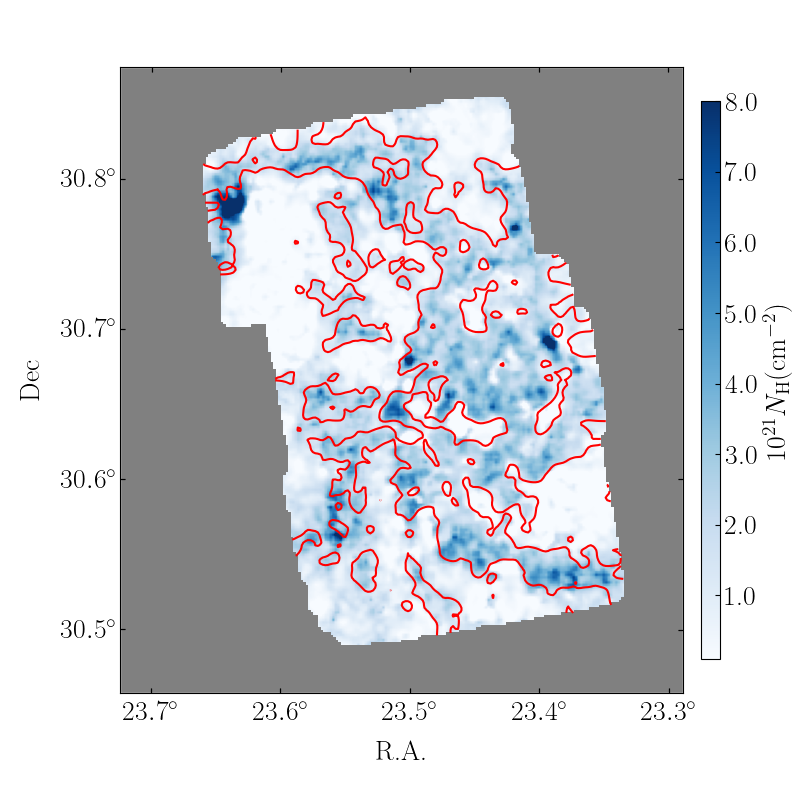}\end{overpic}%
	}
		\end{minipage}\hfill
		\begin{minipage}[t]{0.49\textwidth}
		\centering
		\begin{overpic}[height=8cm]{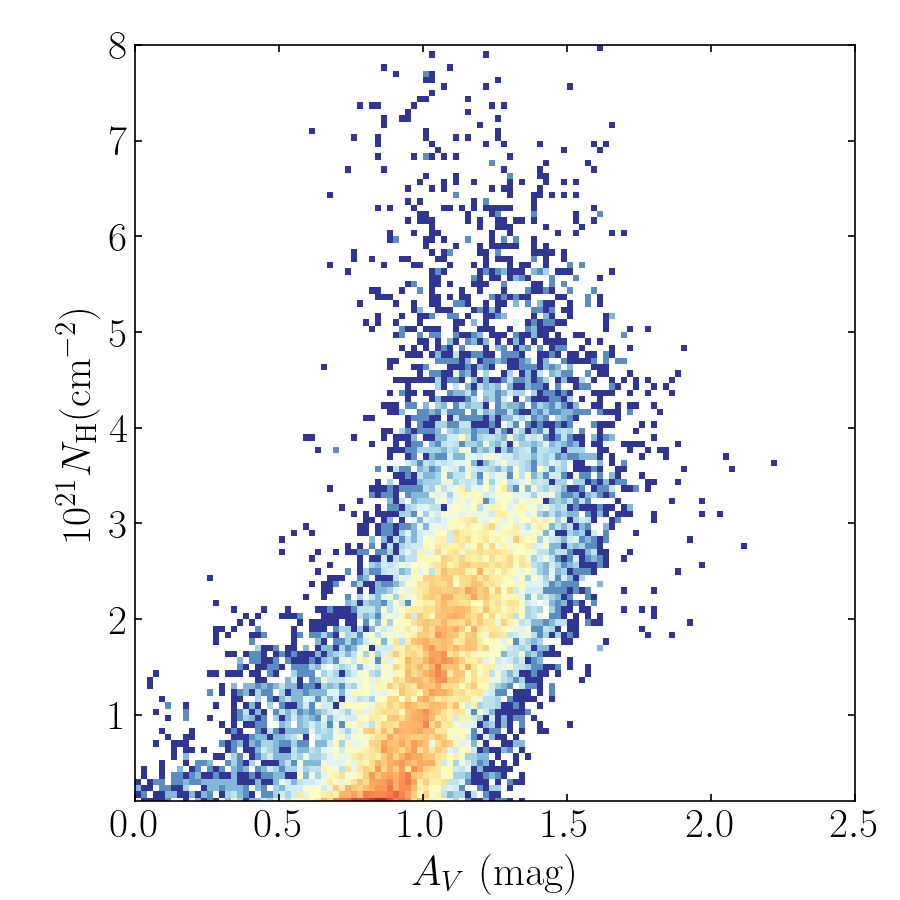}\end{overpic}
		\end{minipage}
	\vspace{0.5em} 
  
	\begin{minipage}[t]{0.49\textwidth}
		\centering
		\raisebox{-4mm}{%
	\begin{overpic}[height=8.9cm]{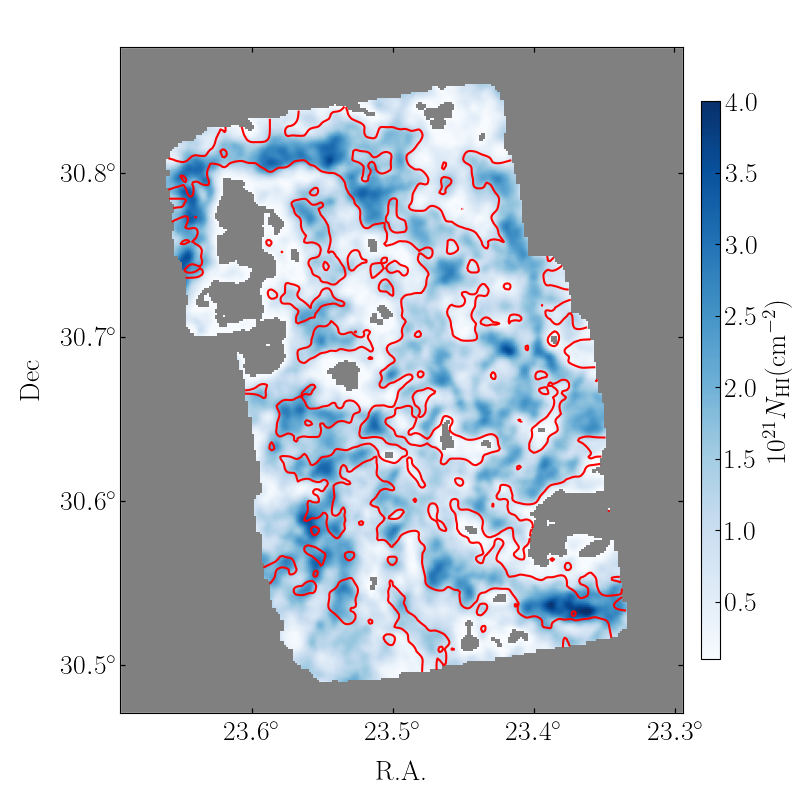}\end{overpic}%
	} 
		\end{minipage}\hfill
		\begin{minipage}[t]{0.49\textwidth}
		\centering
		\begin{overpic}[height=8cm]{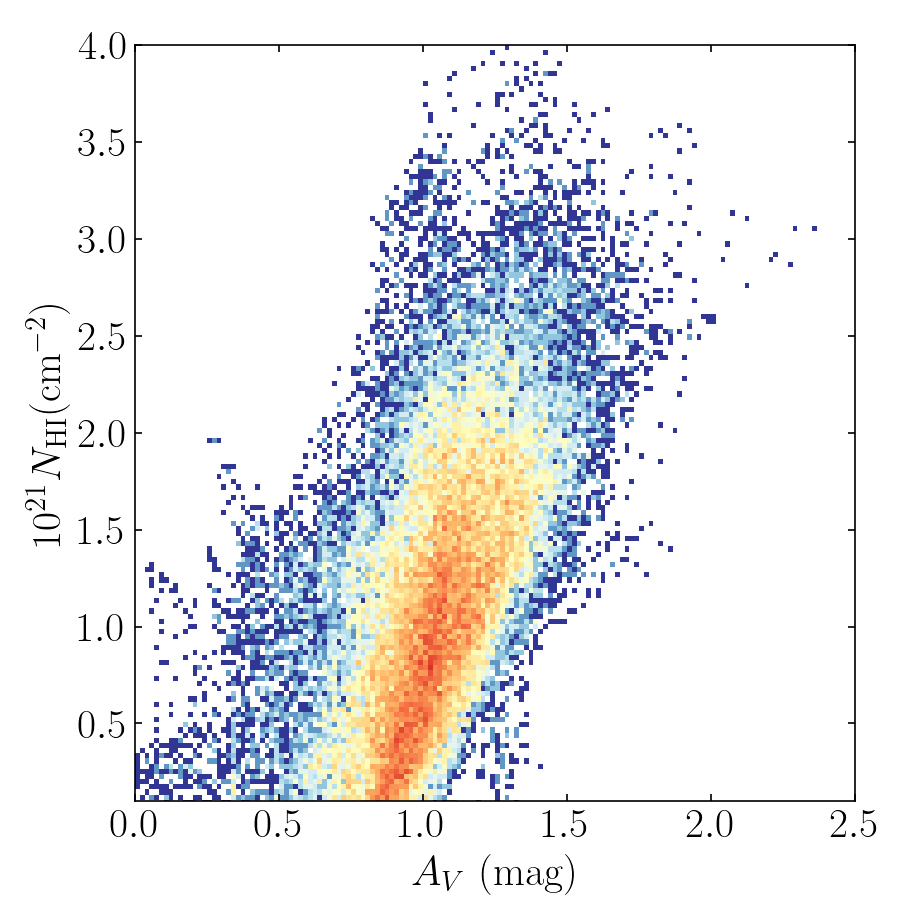}\end{overpic}
		\end{minipage}
		\caption{Left panels: maps of H (\citealt{2024A&A...688A.171K}, upper left), and H I (\citealt{2010A&A...522A...3G}, lower left) column densities, with red contours indicating the mean $A_V$ value as determined in this work.
		Right panels: pixel-by-pixel comparison between extinction values and gas column densities.
		For each corresponding gas component, the horizontal axis shows the extinction value in the $V$ band for each pixel determined in this work and the vertical axis shows the column density. 
		The color scale from blue to red represents the number of pixels in each bin (from low to high).}
	\label{fig:com_H}
  \end{figure}

Although extinction is fundamentally linked to the dust content within galaxies, it is well established that dust and gas are closely associated in the interstellar medium. 
We present the spatial distributions across M33 of the total hydrogen (H) column density with a resolution of 18.2$^{\prime\prime}$ derived from SED fits to \emph{Herschel} maps \citep{2024A&A...688A.171K}, and the neutral atomic hydrogen (H I) column density with a resolution of 17$^{\prime\prime}$ traced by VLA H I observations \citep{2010A&A...522A...3G}.
 As shown in the left panels of Figure \ref{fig:com_H}, the extinction distribution constructed in this work exhibits a morphology very similar to both the H and H I distributions, indicating that the extinction map effectively traces the overall gas structure across M33. 

After convolving the extinction distribution with Gaussian kernels to match the resolutions of the H and H I distributions, we perform a pixel-by-pixel quantitative comparison of the extinction values with both the H and H I column densities, as displayed in the right panels of Figure \ref{fig:com_H}.
As shown in the upper right panel, there is a clear positive correlation between extinction ($A_V$) and the total hydrogen column density ($N_\mathrm{H}$). 
Regions with higher $A_V$ values generally correspond to higher $N_\mathrm{H}$, indicating that the extinction map constructed in this work effectively traces the overall gas distribution within the disk of M33. 
H I represents the primary phase of atomic gas in galactic disks and forms the widespread neutral layer that pervades the galaxy, whereas $A_V$ primarily reflects dust abundance and distribution.
However, on galactic scales, dust and neutral gas are typically well mixed, with dust grains embedded within the gas (e.g., \citealt{2011piim.book.....D}). 
In the lower right panel of Figure \ref{fig:com_H}, $A_V$ and $N_{\rm HI}$ exhibit an overall positive relationship, indicating that the extinction map serves as a robust tracer of the large-scale distribution of neutral atomic gas in M33. 
The correlation weakens at high $A_V$ due to the formation and accumulation of molecular hydrogen (H$_2$): in diffuse H I clouds, H$_2$ is readily photodissociated, and can persist only when molecular gas becomes sufficiently dense for self-shielding. 
At high $A_V$, the substantial molecular component implies that $A_V$ no longer tracks H I alone, increasing the scatter in the $A_V-N_{\rm HI}$ relation.

\begin{figure}[t]
	\centering
	\begin{minipage}[t]{0.49\textwidth}
	\centering
	\raisebox{-4mm}{%
\begin{overpic}[height=8.9cm]{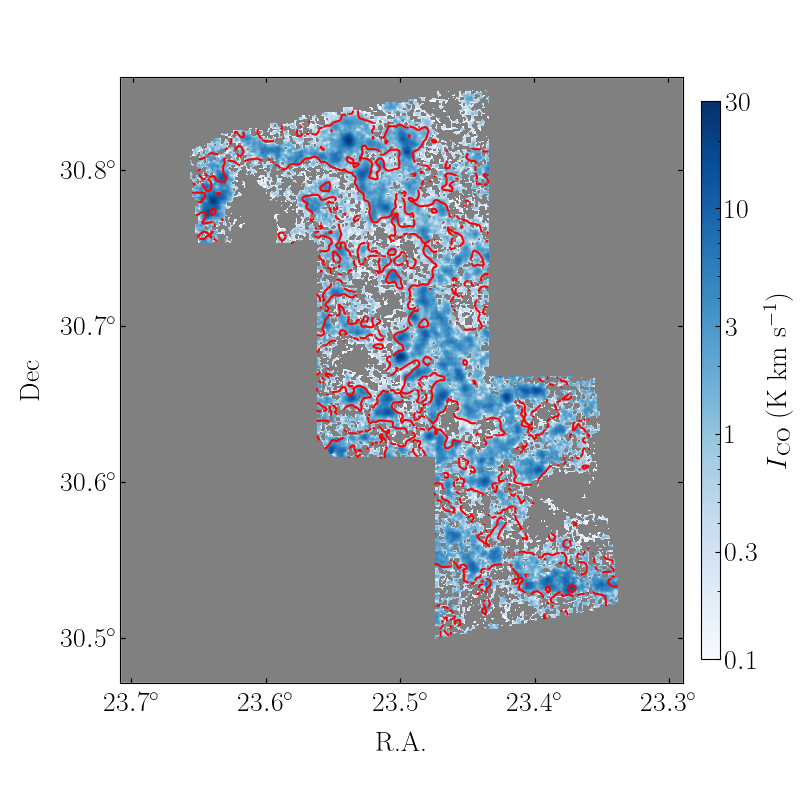}\end{overpic}%
}
	\end{minipage}\hfill
	\begin{minipage}[t]{0.49\textwidth}
	\centering
	\begin{overpic}[height=8cm]{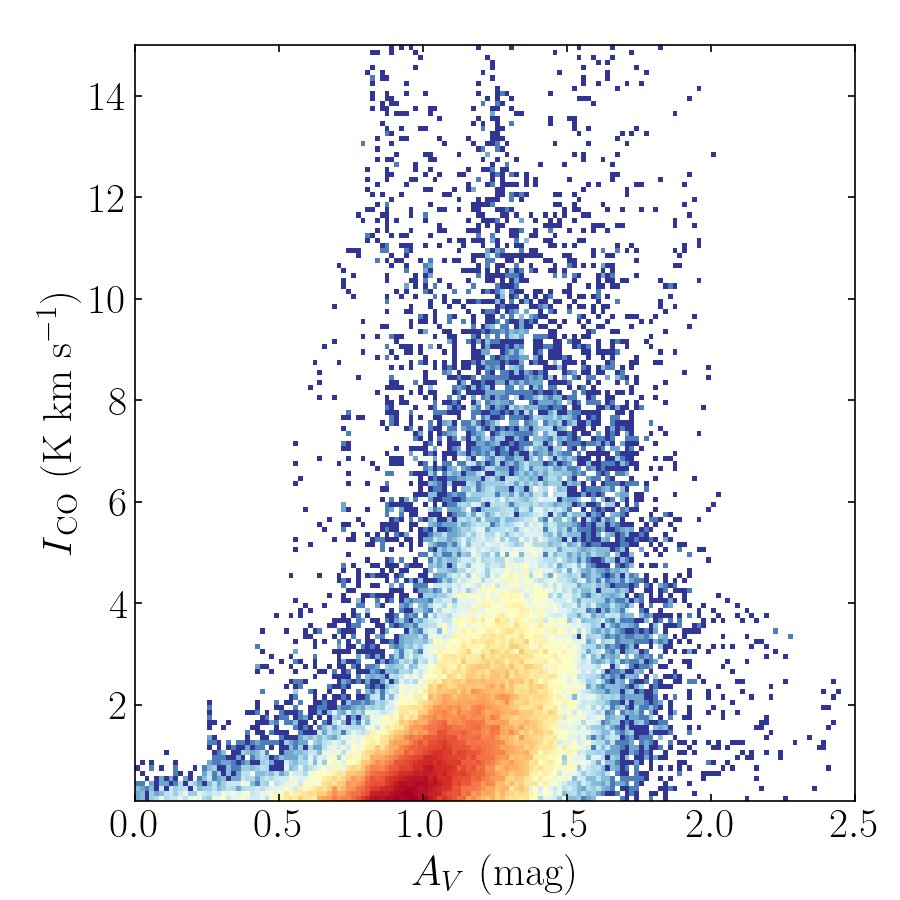}\end{overpic}
	\end{minipage}
	\caption{Left panel: IRAM CO integrated intensity map \citep{2010A&A...522A...3G}, with red contours indicating the mean $A_V$ value as determined in this work.
	Right panel: a two-dimensional density map (density increases from blue to red) showing the relationship between extinction ($A_V$; x-axis) and integrated CO intensity ($I_{\rm CO}$; y-axis). 
	}
	\label{fig:com_CO}
	\end{figure}

In addition to hydrogen, we also compare the extinction map with the integrated CO (2-1) intensity map at 12$^{\prime\prime}$ resolution from observations with the Institut de Radioastronomie Millim\'etrique (IRAM) 30-meter telescope \citep{2010A&A...522A...3G}.
For a pixel-by-pixel quantitative comparison, the extinction distribution derived in this work is convolved to the same resolution as the integrated CO (2-1) intensity map.
As shown in the left panel of Figure \ref{fig:com_CO}, the morphology of the CO integrated intensity distribution closely matches the contours of the average extinction value, indicating that the extinction map constructed in this work effectively traces dense, dust-rich interstellar environments.
The right panel of Figure \ref{fig:com_CO} shows a clear positive, but not strictly linear,  correlation between dust extinction ($A_V$) and integrated CO intensity ($I_{\rm CO}$).
$I_{\rm CO}$ remains low at small $A_V$ and rises rapidly only above an extinction threshold because CO is readily dissociated when shielding is insufficient. 
Only at higher extinction ($A_V \approx$ 1 mag, \citealt{2010ApJ...716.1191W}) does dust provide adequate protection, allowing CO to survive and emit.
In addition, the dispersion at high $A_V$ likely arises as CO emission primarily traces molecular material while large extinction can also originate from non-molecular structures along the sight line, and the $A_V$ values derived in this work are stellar line-of-sight averages that may underrepresent the full column of dust and gas.
Overall, the close spatial association between high CO intensity and large extinction demonstrates that the dust extinction map constructed in this work provides a robust and physically motivated complement to CO emission for probing molecular gas structure and star-forming regions.

\subsection{Comparison with extinction values in previous works}

As shown in the right panel of Figure \ref{fig:M33_extinction_var}, the $A_V$ values derived in this work reach up to 2.5 mag per pixel, consistent with the highest values found in previous works, whereas the average value of approximately 1.05 mag exceeds the reported mean values.
For example, \citet{2022ApJS..260...41W}, who modeled the photometry of OB-type supergiants from UV to near-IR, and \citet{2022AJ....163...16M}, who combined archival images from UV to IR for young star cluster population, both obtained maximum extinction values comparable to that in this work.
However, the mean extinction values in \citeauthor{2022ApJS..260...41W} (\citeyear{2022ApJS..260...41W}, $A_V = 0.43$ mag) and \citeauthor{2022AJ....163...16M} (\citeyear{2022AJ....163...16M}, $E(B-V) < 0.2$ mag for two-thirds of the clusters) are substantially lower, roughly half of the mean value found in this work.
Additionally, \citet{2017PhDT.......221H} divided M33 into 1170 spatial regions and modeled the SEDs from the far-UV to the near-IR to derive extinction values. 
While the derived maximum extinction value $A_V$ also approaches 2.5 mag---similar to the highest extinction values derived in this work, the mean value in \citet{2017PhDT.......221H} remains at 0.53 mag. 
This is about twice the average value reported by \citet{2009A&A...493..453V}, who measured the total attenuation in radial bins in M33 with emissions at various wavelengths (e.g., H$\alpha$, UV and IR) and found a mean $A_V$ value of 0.25 mag.
Overall, the variations in reported extinction values reflect the complex interplay between tracer selection, spatial resolution, and methodology.

A crucial factor underlying the discrepancies is the choice of extinction tracer.
In contrast to OB-type supergiants or young star clusters, we adopt RGB stars as extinction tracers.
RGB stars belong to an older population with a larger vertical scale height and a more diffuse distribution than younger populations (e.g., \citealt{2013A&ARv..21...61R,2016ARA&A..54..529B}), allowing us to trace a thicker portion of the dusty disk along sight lines, rather than being biased toward the near side or the surface layers where young clusters reside. 
As a result, the average $A_V$ value determined in this work naturally exceeds those obtained by methods restricted to younger, less obscured populations.

Spatial resolution and selection effects also play significant roles.
The method adopted in this work, based on individually resolved RGB stars, benefits from high spatial resolution and large sample size, enabling the identification of heavily obscured regions that might be missed or averaged out in lower-resolution or integrated analyses.
Notably, the highest $A_V$ value we derived aligns with the maximum extinction values reported by \citet{2022ApJS..260...41W}, \citet{2022AJ....163...16M}, and \citet{2017PhDT.......221H} toward the most heavily extincted sight lines, suggesting that the extinction distribution constructed in this work is able to capture the broad dynamic range of dust extinction in M33.


\section{Conclusion} \label{sec:conclusion}

Based on individually resolved RGB stars and multiband photometry from the PHATTER survey, we present the first high-resolution extinction distribution across the disk of M33.
Characterized by pronounced spatial variations, the distribution reveals the highly non-uniform distribution of dust, highlighting both large-scale structures and small-scale features within the galaxy.
It also exhibits a strong spatial correspondence with the distributions of total hydrogen, as well as with H I and CO individually, indicating that the extinction map effectively traces both the diffuse and dense components of the interstellar medium in M33.  
The $V$-band extinction per pixel reaches a maximum value of 2.5 mag, in agreement with previous studies, and has a mean value of about 1.05 mag--—significantly higher than earlier results. 
The elevated mean extinction is primarily attributed to the adoption of RGB stars as tracers, which probe the full dust column along the line of sight, as well as the improved spatial resolution of this work. 

Beyond detailing the dust distribution, the extinction map constructed in this work brings into focus the structural features of M33, such as spiral arms, inter-arm regions, and localized dust clouds, and offers new perspectives on how dust is arranged relative to the galactic structure.
Serving as a valuable foundation for accurate extinction correction, the derived map will benefit future studies in M33, including upcoming observations with the Chinese Space Station Telescope; it further contributes to a deeper understanding of the interstellar medium and star formation processes in nearby galaxies.

\acknowledgments 
We are very grateful to the anonymous reviewer for the constructive comments and suggestions, which have significantly improved this work.
It is a pleasure to thank Profs. Biwei Jiang, Haibo Yuan and Di Li for the invaluable discussions and insights. 
This work is supported by the National Natural Science Foundation of China under project 12303030 and the Shandong Provincial Natural Science Foundation under project ZR2024QA236.
Additional support is provided by the National Natural Science Foundation of China (12203025, 12133002, U2031209, 12173034, 12322304), the Shandong Provincial Natural Science Foundation (ZR2022QA064), the CSST Projects (CMS-CSST-2021-A09 and CMS-CSST-2025-A11), and Shandong Provincial University Youth Innovation and Technology Support Program (No. 2022KJ138).
The numerical computations were conducted on the Qilu Normal University High Performance Computing, Jinan Key Laboratory of Astronomical Data.

%





\bibliography{paper}{}
\bibliographystyle{aasjournal}



\end{CJK*}
\end{document}